\documentstyle[preprint,tighten,aps,psfig]{revtex}

\begin{document}
\title{ Restoration of chiral symmetry in excited hadrons.}
\author{ L. Ya. Glozman}
\address{  Institute for Theoretical
Physics, University of Graz, Universit\"atsplatz 5, A-8010
Graz, Austria\footnote{e-mail: leonid.glozman@uni-graz.at}}
\maketitle

\begin{abstract} 
Physics of the low-lying and high-lying hadrons in the
light flavor sector is reviewed. While the low-lying
hadrons are strongly affected by the spontaneous breaking
of chiral symmetry, in the high-lying hadrons the chiral
symmetry is restored. A manifestation of the chiral symmetry
restoration in excited hadrons is a persistence of the
chiral multiplet structure in both baryon and meson
spectra. Meson and baryon chiral multiplets are classified. A
relation between the chiral symmetry restoration and the
string picture of excited hadrons is discussed.
\end{abstract}

\bigskip
\bigskip

\section{Introduction}
It was believed by many people (and is still believed by some)
that there should be some universal physical picture (model) for
all usual hadrons.\footnote{Under "usual" hadrons we assume those ones
which are not glueballs and  with quantum
numbers which are provided by the minimal  $\bar q q$ or
$qqq$ quark Fock component.} 
If we consider, as example, atoms, there is indeed
a universal picture for all excitations: electrons move in the central
Coulomb field of the nucleus. Such a system is essentially nonrelativistic
and relativistic effects appear only as very small corrections to
the nonrelativistic description. However,  hadrons in the $u,d,s$
sector are more complex systems. This complexity comes in particular from
the very small masses of $u$ and $d$ quarks. These small masses
guarantee that the role of relativistic effects, such as creation 
of pairs from the vacuum, should be important. If so  in the
$u,d,s$ quark sector a description should
incorporate valence quarks, sea quarks and gluonic degrees of freedom.\\

The main message of these lectures is that physics of the low-lying
hadrons in the $u,d,s$ sector is essentially different from the
physics of the highly excited states. In the former case
spontaneous breaking of chiral symmetry is crucial for physics
implying such effective degrees of freedom as constituent quarks
(being essentially quasiparticles \cite{NJL}), constituent quark -
Goldstone boson coupling \cite{NJL,MG}, etc. In the latter case,
on the other hand, spontaneous breaking of chiral symmetry in the
QCD vacuum becomes irrelevant, which is referred to as effective
chiral symmetry restoration or chiral symmetry restoration of the 
second kind \cite{G1,CG1,CG2,G2,G3}. Hence in this case other
degrees of freedom become appropriate and probably the string
picture \cite{NAMBU} with "bare" quarks of definite chirality
at the ends of the string \cite{G4} is a relevant description.\\

It is not a surprise that physics of the high-lying excitations and
of the low-lying states is very different in complex systems. 
Remember that in Landau's
Fermi-liquid theory (QCD is a particular case of such a theory)
the quasiparticle degrees of freedom are relevent only to the
low-lying excitations while high-lying levels are excitations
of bare particles.\\

These lectures  consist of the following sections. In the
second one we review chiral symmetry of QCD. The third section
is devoted to a description of the low-lying hadrons, which are
strongly affected by spontaneous breaking of chiral symmetry. 
Empirical hadron spectra are reviewed in the section IV.
In the fifth
section we introduce chiral symmetry restoration in highly
excited hadrons. In the next section a toy pedagogical model
will be discussed which clearly illustrates that there is no
mystery in symmetry restoration in high-lying spectra.
Implications of the quark-hadron duality in QCD for spectroscopy
are discussed in  section VII.
 In  sections
VIII and IX we will classify chiral multiplets of excited mesons and
baryons respectively. In  section X it is shown that a simple
potential constituent quark model is incompatible with the
chiral symmetry restoration in excited hadrons.
 A relation between the chiral symmetry 
restoration and the string picture of excited hadrons is discussed
in  section XI. Finally, a short summary will be presented in the
conclusion part.

\section{Chiral symmetry of QCD}

Consider the chiral limit where quarks are massless. It
is definitely justified for  $u$ and $d$ quarks since
their masses are quite small  compared to $\Lambda_{QCD}$ and
the typical hadronic scale of 1 GeV; in good approximation
they can be neglected. Define the right- and left-handed components
of quark fields

\begin{equation}
\psi_R = \frac{1}{2}\left( 1+\gamma_5 \right ) \psi,~~
\psi_L = \frac{1}{2}\left( 1-\gamma_5 \right ) \psi.
\label{RL}
\end{equation}

\noindent
If there is no interaction, then the right- and left-handed
components of the quark field get decoupled, as it is well
seen from the kinetic energy term

\begin{equation}
{\cal L}_0 = i \bar \Psi \gamma_\mu \partial^\mu \Psi =
i \bar \Psi_L \gamma_\mu \partial^\mu \Psi_L +
i \bar \Psi_R \gamma_\mu \partial^\mu \Psi_R,
\label{L0}
\end{equation}

\noindent
see Fig. 1. 

\begin{figure}
\centerline{
\psfig{file=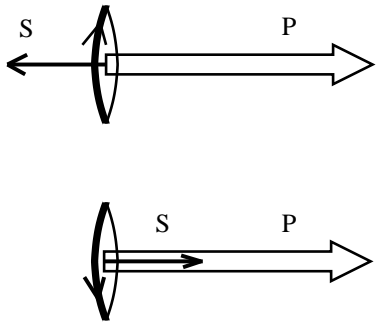,width=0.3\textwidth}
}
\caption{Left-handed and right-handed massless fermions.}
\end{figure}

In QCD the quark-gluon interaction Lagrangian
is vectorial, 
$\bar \psi \gamma^\mu \psi A_\mu$, which does not mix the right- and
left-handed components of quark fields. Hence in the chiral limit
 the left- and right-handed components of
quarks are completely decoupled in the QCD Lagrangian. Then,
assuming only one flavor of quarks such a Lagrangian is invariant
under two independent global variations of phases of the
left-handed and right-handed quarks:

\begin{equation}
\psi_R \rightarrow 
\exp \left( \imath \theta_R \right)\psi_R; ~~
\psi_L \rightarrow 
\exp \left( \imath \theta_L\right)\psi_L.
\label{CH}
\end{equation}

\noindent
Such a transformation can be identically rewritten in terms
of the vectorial and axial transformations: 

\begin{equation}
\psi\rightarrow 
\exp \left( \imath \theta_V \right)\psi; ~~
\psi \rightarrow 
\exp \left( \imath \theta_A  \gamma_5 \right)\psi.
\label{VA}
\end{equation}

\noindent
The symmetry group of these  phase transformations is

\begin{equation}
U(1)_L \times U(1)_R = U(1)_A \times U(1)_V.
\label{VAS}
\end{equation}

Consider now the chiral limit for two flavors, $u$ and $d$. 
The quark-gluon interaction Lagrangian is insensitive
to the specific flavor of quarks.  For example, one can substitute
 the $u$ and $d$ quarks by  properly normalized orthogonal 
linear combinations of $u$ and $d$
quarks ({\it i.e.} one can perform a rotation in the isospin space)
and nothing will change. Since the left- and right-handed components
are completely decoupled, one can perform two independent isospin
rotations of the left- and right-handed components:

\begin{equation}
\psi_R \rightarrow 
\exp \left( \imath \frac{\theta^a_R \tau^a}{2}\right)\psi_R; ~~
\psi_L \rightarrow 
\exp \left( \imath \frac{\theta^a_L\tau^a}{2}\right)\psi_L,
\label{ROT}
\end{equation}

\noindent
where $\tau^a$ are the isospin Pauli matrices and the
angles $\theta^a_R$ and $\theta^a_L$ parameterize rotations
of the right- and left-handed components. These rotations leave
the QCD Lagrangian invariant. The symmetry group of these
transformations,

\begin{equation}
SU(2)_L \times SU(2)_R,
\label{chsymm}
\end{equation}

\noindent
is called chiral symmetry.\\

Actually in this case the Lagrangian is also invariant under
the variation of the common phase of the left-handed $u_L$ and $d_L$ quarks, 
which is the $U(1)_L$ symmetry and similarly - for the right-handed 
quarks. Hence the total chiral symmetry group of the QCD Lagrangian
is

\begin{equation}
U(2)_L \times U(2)_R = SU(2)_L \times SU(2)_R \times U(1)_L \times U(1)_R
= SU(2)_L \times SU(2)_R \times U(1)_V \times U(1)_A.
\label{tot}
\end{equation} 

\noindent
This is a symmetry of the QCD Lagrangian at the classical level. At
the quantum level the $U(1)_A$ symmetry is explicitly broken due to
axial anomaly, which is effect of quantum fluctuations.
The $U(1)_V$ symmetry is responsible for the baryon number conservation
and will not be discussed any longer.\\

Now generally if the Hamiltonian of a system is invariant under
some transformation group $G$, then one can expect
that one can find states which are simultaneously eigenstates
of the Hamiltonian and of the Casimir operators of the group, $C_i$.
Now, if the ground
state of the theory, the vacuum, is invariant under the same
group, {\it i.e.}  if for all $U \in G$
\begin{equation}
 U | 0 \rangle = | 0 \rangle ,
\label{vac}
\label{symvaccond}\end{equation}
then eigenstates of this Hamiltonian corresponding to excitations
above the vacuum can be grouped into degenerate multiplets corresponding
to the particular representations of $G$. 
 This mode of symmetry is
usually referred to as the Wigner-Weyl mode.  Conversely, if 
(\ref{symvaccond}) does not hold,  the excitations do not
generally form degenegerate multiplets in this case.  This situation
is called spontaneous symmetry breaking.\\ 

If chiral symmetry were realized in the Wigner-Weyl mode, then the excitations
would be grouped into representations of the chiral group. 
The representations of the chiral group are discussed in
detail in   the following sections. The important
feature is that the every representation except the trivial one 
 necessarily implies
parity doubling. In other words, for every baryon with
the given quantum numbers and parity, there must exist another
baryon with the same quantum numbers but opposite parity and
which must have the same mass.
In the case of mesons the chiral representations combine, e.g.
the pions with the $f_0$ mesons, which should
be degenerate. This feature is definitely not observed
for the low-lying states in hadron spectra. This means that
Eq.~(\ref{symvaccond}) does not apply;
the continuous chiral symmetry of the QCD Lagrangian is spontaneously (dynamically)
broken in the vacuum, {\it{i.e.}} it is hidden. Such a mode
of symmetry realization is referred to as the Nambu-Goldstone one.
\\

The independent left and right rotations (\ref{ROT}) can be
represented equivalently  with independent isospin and axial 
rotations

\begin{equation}
\psi \rightarrow 
\exp \left( \imath \frac{\theta^a_V \tau^a}{2}\right)\psi; ~~
\psi \rightarrow 
\exp \left( \imath  \gamma_5 \frac{\theta^a_A\tau^a}{2}\right)\psi.
\label{VA}
\end{equation}

\noindent
The existence of 
 approximately degenerate isospin multiplets
in hadron spectra suggests that the vacuum is invariant under
the isospin transformation. Indeed, from the theoretical side the Vafa-Witten
theorem \cite{VW} guarantees that  in the local gauge theories
the vector part of chiral symmetry cannot be spontaneously broken.
The axial  transformation mixes
states with opposite parity. The fact that the low-lying states
do not have parity doublets
implies that the vacuum is not invariant under the
axial  transformations. In other words the almost perfect
chiral symmetry of the QCD Lagrangian is dynamically broken
 by the vacuum down to the vectorial (isospin) subgroup

\begin{equation}
SU(2)_L \times SU(2)_R \rightarrow SU(2)_I.
\label{breaking}
\end{equation}

The noninvariance of the vacuum with respect to the three axial
 transformations requires existence of three massless
Goldstone bosons, which should be pseudoscalars and form an
isospin triplet. These are identified with pions. The nonzero
mass of pions is entirely due to the {\it explicit} chiral symmetry breaking
by the small masses of $u$ and $d$ quarks. These small masses can
be accounted for as a perturbation. As a result the squares of
the pion masses are proportional to the  $u$ and $d$ quark masses
\cite{GOR}

\begin{equation}
m_{\pi}^2 = -\frac{1}{f_{\pi}^2} \frac{m_u + m_d}{2} 
(\langle \bar u u \rangle + \langle \bar d d \rangle) + O (m_{u,d}^2).
\label{gor1}
\end{equation}

That the vacuum is not invariant under the axial transformation
is directly seen from the nonzero values of the quark condensates,
which are  order parameters for spontaneous chiral
symmetry breaking. These condensates are the vacuum expectation
values of the $\bar \psi \psi = \bar \psi_L \psi_R + \bar \psi_R \psi_L$
operator and at the renormalization scale of 1 GeV they approximately are

\begin{equation}
\langle \bar u u \rangle  \simeq \langle \bar d d \rangle  \simeq -
 (240 \pm 10 MeV)^3.
\label{con}
\end{equation}

\noindent
The values above are deduced from phenomenological considerations \cite{L}.
Lattice
gauge calculations also confirm the nonzero and rather large values
for quark condensates. However, the quark condensates above are not
the only order parameters for chiral symmetry breaking. There exist
chiral condensates of higher dimension (vacuum expectation values of
more complicated combinations of  $\bar \psi$  and $ \psi$ that are
not invariant under the axial transformations). Their numerical values
are difficult to extract from phenomenological data, however, and
they are still unknown.\\

To summarize this section. There exists overwhelming evidence that
the nearly perfect chiral symmetry of the QCD Lagrangian is
spontaneously broken in the QCD vacuum. Physically this is because
 the vacuum state in QCD is highly nontrivial which can be seen
by the condensation in the vacuum state of the chiral pairs. These condensates
break the symmetry of the vacuum with respect to the axial transformations
and as a consequence, there is no parity doubling in the low-lying spectrum.
However, as we shall show, the role of the
chiral symmetry breaking quark condensates becomes progressively 
less important 
once we go up in the spectrum, {\it i.e.} the chiral symmetry is
effectively restored, which should be evidenced by the systematical appearance
of the approximate parity doublets in the highly lying spectrum. This is the
subject of the following sections.

\section{A few words about chiral symmetry breaking and
low-lying hadrons}

A key to understanding of the low-lying hadrons is
spontaneous breaking of chiral symmetry (SBCS). Hence it is
instructive to overview physics of SBCS. An insight into
this phenomenon is best obtained from the pre-QCD
Nambu and Jona-Lasinio model \cite{NJL}. Its application to
such questions as formation of constituent quarks as
quasiparticles in the Bogoliubov sense, their connection to the
quark condensate and appearance of the low-lying collective
excitations - Goldstone bosons - was a subject of intensive research
for the last two decades and is reviewed e.g. in Ref. \cite{VW}.
Actually all microscopical models of SBCS in QCD, such as
based on instantons  \cite{SD} or other topological configurations,
or on nonperturbative resummation of gluon exchanges \cite{A}, or on
assumption that the Lorentz scalar confining interaction is
an origin for SBCS \cite{SIMONOV}, all share the key elements and
ideas of the NJL picture. The only essential difference between
all these models is a specification of those  interactions
that are responsible for SBCS.\\

Any interquark interaction in QCD mediated by the
intermediate gluon field, in the local approximation,
contains as a part a chiral-invariant 4-fermion interaction

\begin{equation}
(\bar \psi \psi)^2 + (\bar \psi i \gamma_5  \vec \tau \psi)^2.
\label{njl}
\end{equation}

\noindent
The first term represents Lorentz-scalar interaction. This
interaction is an attraction between the left-handed
quarks and the right-handed antiquarks and vice versa. When it
is treated nonperturbatively in the mean-field approximation,
which is well justified in the vacuum state, it leads to the
condensation of the chiral pairs in the vacuum state

\begin{equation}
\langle 0 | \bar \psi \psi | 0 \rangle = 
\langle 0 | \bar \psi_L \psi_R +  \bar \psi_R \psi_L | 0 \rangle  \neq 0 .
\label{con}
\end{equation}

\noindent
Hence it breaks chiral symmetry, which is a
nonperturbative phenomenon. This dynamics is described
by the famous gap equation which is similar to the one
of Bardeen-Cooper-Schrieffer theory of superconductivity.
This attractive interaction
between bare quarks can be  absorbed into a mass
of a quasiparticle. This is provided by means of Bogoliubov
transformation: Instead of operating with the original bare
quarks and antiquarks one introduces quasiparticles. Each
quasiparticle is a coherent superposition of bare quarks
and antiquarks. Bare particles have both well-defined helicity
and chirality,
while quasiparticles have only definite helicity and contain
a mixture of bare quarks and antiquarks with opposite chirality.
This trick allows us to absorb the initial Lorentz-scalar
attractive interaction between the bare quarks into a mass of the
quasiparticles. These quasiparticles with dynamical mass can
be associated with the constituent quarks. An important feature
is that this dynamical mass appears only at low momenta, below
the ultraviolet cutoff  $\Lambda$ in the NJL model, i.e. where
the low-momentum attractive interaction between quarks is
operative. All quarks with momenta higher than $\Lambda$ remain
undressed. In reality, of course, this step-function behaviour 
of the dynamical mass should be substituted by some smooth function.
Hence in the vaccum a system of massless interacting
quarks  at low momenta can be effectively substituted by a system
of the {\it noninteracting} quasiparticles with dynamical mass $M$.
This mechanism of dynamical symmetry breaking and of
 creation of quasiparticles with dynamical
mass  is a very general one and persists
in different many-fermion systems - from  the superconductors to 
the atomic nuclei.\\

\begin{figure}
\centerline{
\psfig{file=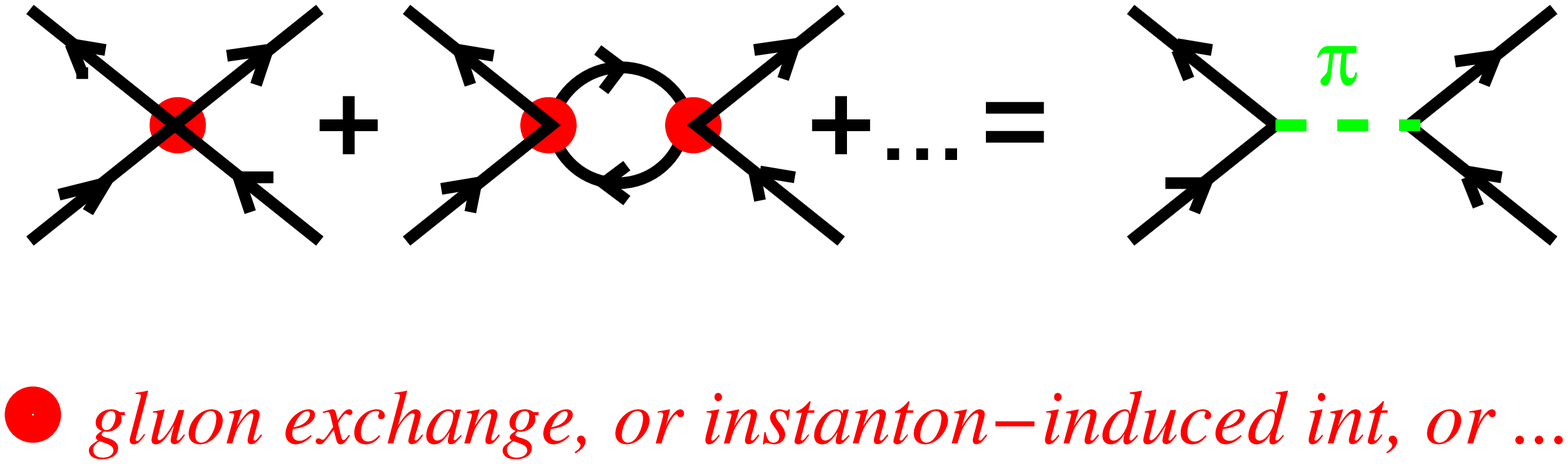,width=0.7\textwidth}
}
\caption{Pion as a relativistic bound state in the quark-antiquark
system.}
\end{figure}

Once the chiral symmetry is spontaneously broken, then there
must appear collective massless Goldstone excitations. Microscopically
their zero mass is provided by the second term of Eq. (\ref{njl}).
This term represents an attraction between the constituent quark and
the antiquark with the pion quantum numbers. Without this term
the pion would have a mass of $2M$. When this term is nonperturbatively
and relativistically iterated, see Fig. 2, the attraction between the constituent quarks
in the pion exactly compensates the $2M$ energy and the pion becomes
massless. This happens because of the underlying chiral symmetry
since it is this symmetry dictates that the strengths of the interactions
represented by the first and by the second terms in Eq. (\ref{njl})
are equal. So the pion is a relativistic bound state of two quasiparticles.
It contains $\bar Q Q, \bar Q Q \bar Q Q, ...$ Fock components. The pion
(as any Goldstone boson) is a highly collective excitation in terms
of the original (bare) quarks and antiquarks  $q$ and $\bar q$ because
the quasiparticles $Q$ and $\bar Q$  themselves are coherent collective 
excitations of bare quarks.\\

\begin{figure}
\centerline{
\psfig{file=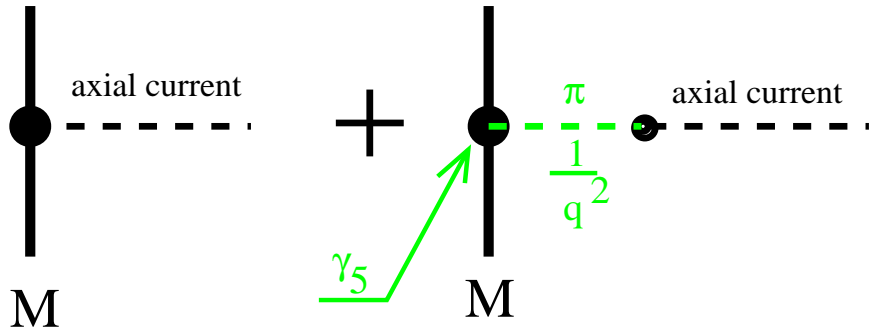,width=0.7\textwidth}
}
\caption{A full axial current in the symmetry broken regime.}
\end{figure}

Now we will go to the low-lying baryons. A basic ingredient of the
chiral quark picture of Manohar and Georgi \cite{MG} is that
the constituent quarks inside the nucleon are strongly coupled
to the pion field and this coupling is regulated by the
Goldberger-Treiman relation. Why this should be so can be seen directly
from the Nambu and Jona-Lasinio mechanism of chiral symmetry breaking.
In terms of the massless bare quarks the axial current,
$A_\mu = \bar \psi \gamma_\mu  \gamma_5 \vec \tau \psi$, is
conserved, $\partial^\mu A_\mu =0$. If one works in terms of free massive
quasiparticles, then it is not conserved, $\partial^\mu A_\mu =2i M 
\bar \psi  \gamma_5 \vec \tau \psi$. How to reconcile this? The only
solution is that the full axial current in the symmetry broken
regime (which must be conserved) contains in addition a term
which exactly cancels $ 2i M \bar \psi  \gamma_5 \vec \tau \psi$.
It is straightforward to see that this additional term must
represent a process where the axial current creates from the vacuum a massless
pseudoscalar isovector boson and this boson in turn couples to the
quasiparticle, see Fig. 3. It is this consideration which forced Nambu
to postulate in 1960 an existence of the Nambu-Goldstone boson
in the symmetry broken regime, which must be strongly coupled
to the quasiparticle.\\

It was suggested in Ref. \cite{GR} that in the low-momentum
regime (which is responsible for masses) the low-lying
baryons in the $u,d,s$ sector can be approximated as systems
of three confined constituent quarks with the residual interaction
mediated by the Goldtone boson field. Such a model was designed
to solve a problem of the low-lying baryon spectroscopy. Microscopically
this residual interaction appears from the t-channel iterations
of those gluonic interactions in QCD which are responsible for
chiral symmetry breaking \cite{GV}, see Fig. 2. An essential feature
of this residual interaction is that it is a flavor- and spin-exchange
interaction of the form

$$ - flavor(i) \cdot flavor(j)  ~spin(i) \cdot spin(j). $$

\noindent
This specific form of the residual interaction between valence
 constituent quarks in baryons allows us  not only to generate
octet-decuplet splittings but what is more important to solve
at the same time
the long-standing puzzle of the ordering of the lowest excitations
of positive and negative parity in the $u,d,s$ sector. This physics
is  a subject of intensive lattice studies and recent
results \cite{KENTUKKY,SASAKI,BGR,SIMULA} do show that the correct
ordering is achieved only close to the chiral limit and hence 
is related to spontaneous breaking of chiral symmetry. 
The  results  \cite{BGR} also
evidence a node in the wave function of the
 radial excitation  of the nucleon (Roper resonance)
 which is consistent with the 3Q leading Fock component
of this state.

\section{Low- and high-lying hadron spectra}

If one looks carefully at the nucleon excitation spectrum,
see Fig. 4, one immediately notices regularities for the
high-lying states starting approximately from the $M \sim 1.7$
GeV region. Namely the nucleon (and delta) high-lying states
show obvious patterns of parity doubling: The states of the
same spin but opposite parity are approximately degenerate. There
are couple of examples where such parity partners have not yet
experimentally been seen. Such doublets are definitely absent
in the low-lying spectrum.
 The high-lying hadron spectroscopy
is a difficult experimental task and the high-lying spectra
 have never been systematically
explored. However, it is conceptually important to answer a question
whether the parity partners exist systematically or not. If yes,
and the existing data hint at it, then it would mean that some
symmetry should be behind this parity doubling and this symmetry
is not operative in the low-lying spectrum. What is this symmetry
and why is it active only in the high-lying part of the spectrum?
Clearly, if the parity doubling is systematic, then it rules
out a description of the highly-excited states in terms of the
constituent quarks (it will be discussed in one the following
sections). Hence the physics of the low-lying
and high-lying states is very different. \\

\begin{figure}
\centerline{
\psfig{file=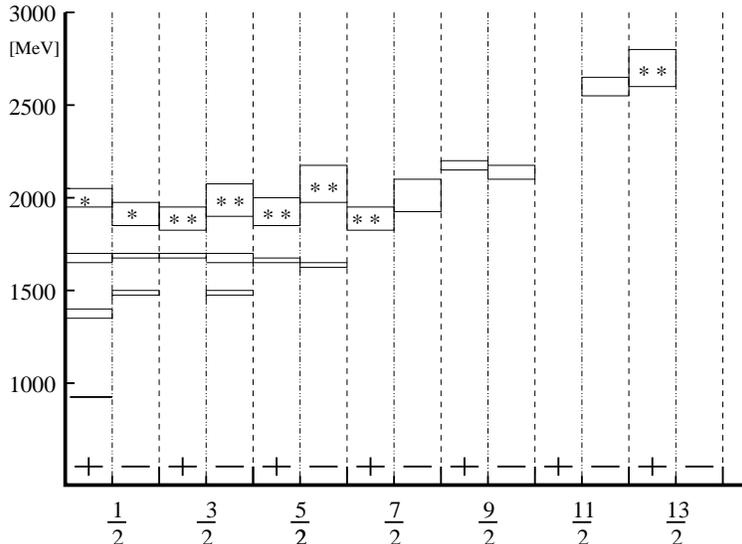,angle=-90,width=0.6\textwidth}
}
\caption{Nucleon excitation spectrum. Those states which
are not yet established are marked by ** or * signs according
to PDG classification.}
\end{figure}

It has been suggested some time ago that  this parity doubling
reflects  restoration of the spontaneously broken
chiral symmetry of QCD \cite{G1}.
We have already discussed in the previous sections that the
underlying chiral symmetry of the QCD Lagrangian would imply,
if the QCD vacuum was trivial, a systematical parity doubling
through the whole spectrum.
However, the chiral symmetry of QCD is dynamically broken
in the QCD vacuum, which leads to the appearance of the
constituent quarks. The constituent (dynamical) mass of quarks
results from their coupling to the quark condensates of the
vacuum.
 We have also discussed that a description
in terms of the constituent quarks makes sense only at low 
momenta. Typical momenta of valence quarks in the low-lying
hadrons are below the chiral symmetry breaking scale, hence
the chiral symmetry is broken in the low-lying states.
The idea of Ref. \cite{G1} was that the typical momenta 
of valence quarks in highly excited hadrons are higher than
the chiral symmetry breaking scale and hence these valence
quarks decouple from the quark condensates of the QCD vacuum.
Consequently the chiral symmetry is effectively restored in highly
excited hadrons.\\

Clearly, if the chiral symmetry restoration  indeed occurs,
then it must be seen also in excited mesons. There are no
systematic data on highly excited mesons in PDG. If one uses
results of the recent systematic partial wave analysis of the proton-antiproton
annihilation at LEAR at 1.8 -2.4 GeV, performed  by the 
London-S.Petersburg group \cite{BUGG1,BUGG2}, then
 once
a careful chiral classification of the states is done
\cite{G2,G3} 
 one clearly
sees direct signs of chiral symmetry restoration,
see, e.g., Fig. 5 where $\pi$ and 
$\bar n n = \frac{\bar u u + \bar d d}{\sqrt 2}$ $f_0$ states are shown
(which must be chiral partners in the chiral symmetry restored
regime). These facts force us to take seriously
the possibility of chiral symmetry restoration in excited hadrons
and also to concentrate  experimental efforts on the
systematical study of highly excited hadrons. Clearly the results
on meson spectroscopy from
 the $\bar p p$ annihilation at LEAR as well as on highly
excited baryons must be checked and {\it completed} at the future facilities
like PANDA at GSI as well as at JPARC and  
at the existing accelerators like at 
JLAB, Bonn, SPRING8, BES, etc. This should be one of the priority tasks.\\

\begin{figure}
\centerline{
\psfig{file=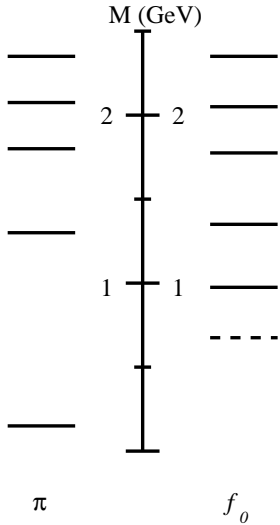,angle=-90,width=0.6\textwidth}
}
\caption{Pion and $\bar n n = \frac{\bar u u + \bar d d}{\sqrt 2}$ $f_0$
states.}
\end{figure}

\section{Chiral symmetry restoration in excited hadrons by definition}

The systematic approach to the symmetry restoration based on
QCD has been formulated in ref. \cite{CG1,CG2}. By definition
an effective symmetry restoration means the following. In QCD
the hadrons with the quantum numbers $\alpha$ are created when
one applies the local interpolating field (current) $J_\alpha$
with such quantum numbers on the vacuum $|0\rangle$. 
This interpolating field contains a combination of valence
quark creation operators at some point $x$.
Then all
the hadrons that are created by the given interpolator appear
as intermediate states in the two-point correlator, see Fig. 6,

\begin{figure}
\centerline{
\psfig{file=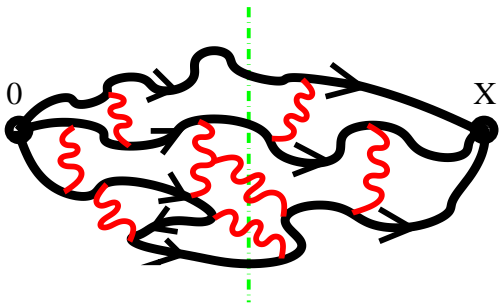,width=0.4\textwidth}
}
\caption{Two-point correlator.}
\end{figure}

\begin{equation}
\Pi =\imath \int d^4x ~e^{\imath q x}
\langle 0 | T \{ J_\alpha (x) J^\dagger_\alpha (0) \} |0\rangle,
\label{corr}
\end{equation}

\noindent
where all possible Lorentz and Dirac indices (specific for
a given interpolating field) have been omitted. Consider
two local interpolating fields $J_1(x)$ and $J_2(x)$ which
are connected by chiral transformation, 

\begin{equation}
J_1(x) = UJ_2(x)U^\dagger,
\label{tr}
\end{equation}
where $U$ is an element of the chiral group. Then, if the vacuum
was invariant under chiral group, $$U|0\rangle = |0\rangle,$$
it follows from (\ref{corr}) that the spectra created by
the operators $J_1(x)$ and $J_2(x)$ would be identical. We know
that in QCD one finds $$U|0\rangle \neq |0\rangle.$$ As a consequence
the spectra of two operators must be in general different. However, it may
happen that the noninvariance of the vacuum becomes unimportant
(irrelevant) high in the spectrum. Then the spectra of both
operators become close al large masses (and asymptotically
identical). This would mean that chiral symmetry is effectively
restored. We stress that this effective chiral symmetry
restoration does not mean that chiral symmetry breaking in
the vacuum disappears, but only that the role of the quark
condensates that break chiral symmetry in the vacuum becomes
progressively less important high in the spectrum \cite{CG1,CG2}.
One could say, that the valence quarks in high-lying
hadrons decouple from the QCD vacuum.
In order to avoid a confusion with the chiral symmetry
restoration in the vacuum state at high temperature or density
one also refers this phenomenon as chiral symmetry restoration
of the second kind.\\

\section{A simple pedagogical example}

It is instructive to consider a very simple quantum
mechanical example of symmetry restoration high in
the spectrum. Though there are conceptual differences
between the field theory with spontaneous symmetry
breaking and the one-particle quantum mechanics (where
only explicit symmetry breaking is possible), nevertheless
this simple example illustrates how this general phenomenon
comes about.

The example we consider is a two-dimensional harmonic
oscillator. We choose the harmonic oscillator only
for simplicity; the property that will be discussed below
is quite general one and can be seen in other systems.
The Hamiltonian of the system is invariant under 
$U(2) = SU(2)\times U(1)$ transformations. This
symmetry has profound consequences on the spectrum of the system.
 The energy levels of this  system are trivially found and
  are given by
\begin{equation}
E_{N, m} \, = \, ( N \, + \, 1 ); ~ m \, =
\, N, N-2, N-4, \, \cdots \, , -(N-2) , -N \; ,
\label{hoeigen}\end{equation}
where $N$ is the principal quantum number  and m is the
(two dimensional) angular momentum.  As a consequence of the symmetry, 
the levels are $(N+1)$-fold
degenerate.

Now suppose we add to the Hamiltonian a $SU(2)$ symmetry breaking
interaction (but which is still $U(1)$ invariant) of the form

\begin{equation}
V_{\rm SB} \, = \, A \, \theta (r - R),
\label{vsb}
 \end{equation}

\noindent
where  $A$ and $R$ are parameters and $\theta$ is the step
function.  Clearly, $V_{\rm SB}$ is not invariant under the
$SU(2)$ transformation. Thus the $SU(2)$ symmetry
is explicitly broken by this additional interaction, that acts
only within a circle of radius $R$.
As a result one would expect that the eigenenergies will not
reflect the degeneracy structure of seen in Eq.~(\ref{hoeigen})
 if the
coefficients $R,A$ are sufficiently large.  Indeed,
 we have solved
numerically for the eigenstates for the case of $A=4$ and $R=1$
in  dimensionless units and one does not see a
multiplet structure in the low-lying spectrum as can be seen in
Fig.~7.

What is interesting for the present context is the high-lying spectrum.
  In Fig.~7 we have also plotted the energies between 70 and 74 for
   a few of the lower $m$'s.
   A multiplet structure is quite evident---to very good approximation
    the states of different $m$'s form degenerate multiplets and,
    although we have not shown this in the figure these multiplets
     extend in $m$ up to $m=N$.  
     
     How does this happen? The symmetry breaking
      interaction  plays a dominant role in the
       spectroscopy for small energies. Indeed, at small
       excitation energies
       the system is mostly located at distances where the symmetry
       breaking interaction acts and where it is dominant.
Hence  the low-lying spectrum to a very large extent is motivated  
by the symmetry breaking interaction.   However, at high 
excitation energies
the system mostly lives at large distances, where physics is dictated
by the unperturbed harmonic oscillator.  Hence at
 higher energies the spectroscopy
       reveals the $SU(2)$ symmetry of the two-dimensional harmonic
        oscillator.

\begin{figure}
\hspace*{-0.5cm}
$\begin{array}{cc}\psfig{file=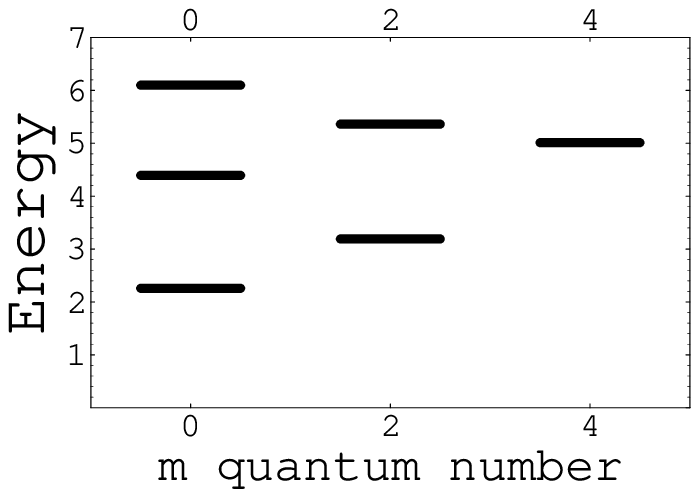,width=0.5\textwidth}
&\hspace*{-0.5cm}\psfig{file=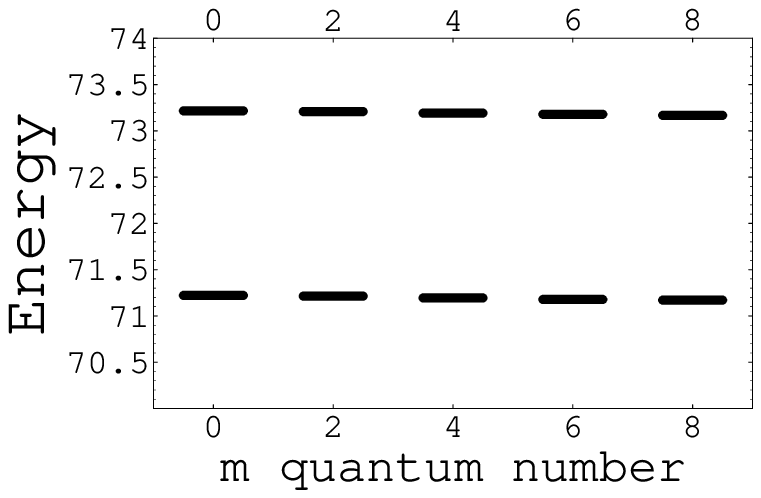,width=0.5\textwidth}\end{array}$
\caption{The low-lying (left panel) and highly-lying (right panel)
spectra of two-dimensional harmonic oscillator with the 
$SU(2)$-breaking term.}
\end{figure}

\section{The quark-hadron duality and chiral symmetry restoration}

A question arises to which extent the chiral symmetry restoration
of the second kind can be theoretically predicted in QCD. There is
a heuristic argument that supports this idea \cite{CG1,CG2}. The
argument is based on the well controlled behaviour of the two-point
function (\ref{corr}) at the large space-like momenta $Q^2 =-q^2$,
where the operator product expansion (OPE) is valid and where all
nonperturbative effects can be absorbed into condensates of
different dimensions \cite{SVZ}. The key point is that all
nonperturbative effects of spontaneous breaking of chiral
symmetry at large $Q^2$ are absorbed into quark condensate
$\langle \bar q q \rangle$ and other quark condensates of
higher dimension. However, the contribution of these
condensates into correlation function is regulated by
the Wilson coefficients. The latter ones are proportional
to $(1/Q^2)^n$, where the index $n$ is determined by the
quantum numbers of the current $J$ and by the dimension
of the given quark condensate. The higher dimension, the larger
$n$. It is important that contributions
of all possible chiral noninvariant
terms of OPE are suppressed by inverse powers of $Q^2$,
the higher dimension of the condensate, the less important
is the given condensate at large $Q^2$. Hence, at large
enough $Q^2$ the two-point correlator becomes approximately
chirally symmetric. At these high $Q^2$ a matching with
the perturbative QCD (where no SBCS) can be done.
In other words, though the chiral symmetry is broken in the
vacuum and all chiral noninvariant condensates are not zero,
their influence on the correlator at asymptotically high $Q^2$
vanishes. This is in contrast to the situation of low
values of $Q^2$, where the role of chiral symmetry breaking
in the vacuum is crucial. Hence, at $Q^2 \rightarrow \infty$ one has

\begin{equation}
\Pi_{J_1}(Q^2) - \Pi_{J_2}(Q^2) \sim \frac{1}{Q^n}, \; \; n>0 \; ,
\label{highQ}
\end{equation}

\noindent
where $J_1$ and $J_2$ are interpolators which are connected
by the chiral transformation according to (\ref{tr}).\\

Now we can use causality of the local field theory and
hence analyticity of the two-point function. Then we can invoke
into analysis a dispersion relation,

\begin{equation}
\Pi_J(Q^2) \, = \,  \int {\rm d}s \frac{\rho_J(s)}{Q^2 + s - i \epsilon},
\label{kl}
\end{equation}
where the spectral density $\rho_J(s)$ is defined as
\begin{equation}
\rho_J (s) \equiv \frac{1}{\pi} \rm{Im} \left ( \Pi_j(s) \right ).
\label{rhoJ}
\end{equation}
The integration in this equation is performed along the cut in Fig. 8.
Since the large $Q^2$ asymptotics of the correlator 
is given by the leading term of
the  perturbation theory,
then the asymptotics of $\rho(s)$ at $s \rightarrow \infty$ must
also be given by the same term of the perturbation theory if the
spectral density approaches a constant value (if it oscillates, then
it must oscillate around the perturbation theory value). Hence
both spectral densities $\rho_{J_1}(s)$ and $\rho_{J_2}(s)$ at
$s \rightarrow \infty$ must approach the same value and the spectral
function becomes chirally symmetric. This theoretical expectation,
that the high $s$ asymptotics of the spectral function is well described
by the leading term of the perturbation theory has been tested e.g.
in the process $e^+e^- \rightarrow hadrons$ , where the interpolator
is given by the usual electromagnetic vector current. This process
is described in standard texts on QCD, for the recent data see
\cite{ee}. Similarly, the vector and the axial vector
spectral densities must coincide in the chiral symmetry restored regime.
They have been measured in the $\tau$ decay by the ALEPH and OPAL
collaborations at CERN \cite{ALEPH,OPAL}. It is well seen from the results
that while the difference between both spectral densities is
large at the masses of $\rho(770)$ and $a_1(1260)$, it becomes
 strongly reduced towards  
$m=\sqrt s \sim 1.7$ GeV.\\

While the argument above about chiral symmetry restoration in
the spectral density is rather general and can be believed to be
experimentally established, strictly speaking it does not necessarily
imply that the high lying hadron resonances must form chiral
multiplets. The reason is that the approximate equility of two spectral
densities would necessarily imply hadron chiral multiplets only if
the spectrum was discrete. In reality, however, the high-lying
hadrons are rather wide overlapping resonances. In addition,
it is only completely continuous non-resonant spectrum that
is described by the chiral invariant leading term of perturbation
theory. Nevertheless, it is indeed reasonable to assume that the
spectrum is still quasidiscrete in the transition region $\sqrt s 
\geq 1.7$ GeV where one approaches the chiral invariant regime.
If so in this region the observed hadrons should
 fall into approximate chiral multiplets.\\

The question arises then what is the functional behaviour that
determines approaching the chiral-invariant regime at large $s$?
Naively one would expect that the operator product expansion
of the two-point correlator, which is valid in the deep
Euclidean domain, could help us. This is not
so, however, for two reasons. First of all, we know
phenomenologically only the lowest dimension quark condensate.
Even though this condensate dominates as a chiral symmetry
breaking measure at the very large space-like $Q^2$, at
smaller $Q^2$ the higher dimensional condensates, which
are suppressed by inverse powers of $Q^2$, are also  
important. These condensates are not known, unfortunately.
But even if we knew all  quark condensates
up to a rather high dimension, it would not help us. This is
because the OPE is only an asymptotic expansion \cite{Z}.
While such kind of expansion is very useful in the space-like
region, it does not define any analytical solution which could
be continued to the time-like region at finite $s$. While
convergence of the OPE can be improved by means of the Borel
transform and it makes it useful for SVZ sum rules for the
low-lying hadrons, this cannot be done for the higher states.
So in order to estimate chiral symmetry restoration effects
 one indeed needs a microscopic theory that would incorporate
{\it at the same time} chiral symmetry breaking and confinement.\\
 
\begin{figure}
\centerline{
\psfig{file=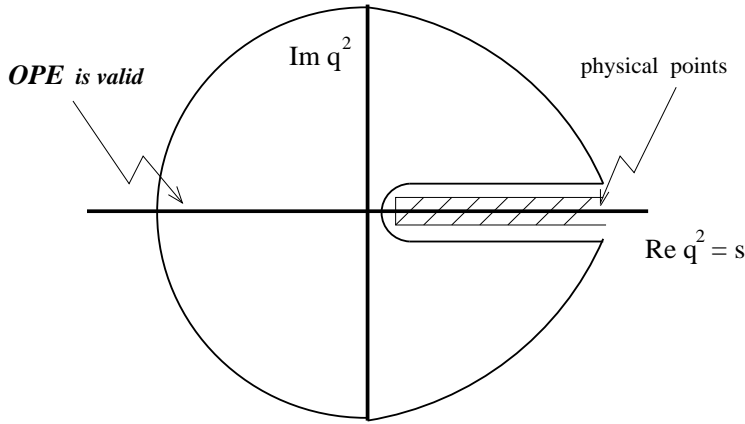,width=0.6\textwidth}
}
\caption{The two-point correlator in the complex $q^2$ plain.}
\end{figure}

\section{Chiral multiplets of excited mesons}

Here we limit ourselves to the two-flavor version of QCD. There
are two reasons for doing this. First of all, the $u$ and $d$ quark
masses are very small as compared to $\Lambda_{QCD}$. Thus the
chiral $SU(2)_L\times SU(2)_R$ and more generally the 
$U(2)_L \times U(2)_R$ symmetries of the QCD Lagrangian are
nearly perfect. This is not the case if the $s$ quark is included,
and a priori it is not clear whether one should regard this
quark as light or "heavy". The second reason is a practical
one -- there are good new data on highly excited $u,d$ mesons
observed in $\bar p p$ annihilation \cite{BUGG1,BUGG2},
but such data are still missing for the strange mesons. Certainly
it would be very interesting and important to extend the analysis
to the $U(3)_L \times U(3)_R$ case. One hopes that the present
results will stimulate the experimental and theoretical
activity in this direction.\\

Mesons reported in Ref. \cite{BUGG1,BUGG2} are obtained
in $\bar p p$ annihilations,
 hence according to OZI rule we have to expect them
to be $\bar q q$ states with $u$ and $d$ valence quark content. Hence
we will    consider  

\begin{equation}
U(2)_L \times U(2)_R = SU(2)_L\times SU(2)_R \times U(1)_V\times U(1)_A, 
\label{sym}
\end{equation}

\noindent
the full chiral group of the QCD Lagrangian. In the following
 chiral symmetry  will refer to specifically the $SU(2)_L\times SU(2)_R$
symmetry. \\

The irreducible representations of 
this group can be specified by the isospins of the left-handed and
right-handed quarks, $(I_L,I_R)$. The total isospin of the state
can be obtained from the left- and right-handed isospins according to
the standard angular momentum addition rules

\begin{equation}
I= |I_L-I_R|, ... , I_L+I_R.
\label{isospin}
\end{equation}

All hadronic states are characterised by a definite parity. However,
not all  irreducible representations of the chiral group are
invariant under parity. Indeed,  parity transforms the left-handed
quarks into the right-handed ones and vice versa. Hence while representations
with $I_L=I_R$ are invariant under parity (i.e. under parity
operation every state in the representation transforms into the
state of opposite parity within the same representation), this
is not true for the case $I_L \neq I_R$. In the latter case 
parity transforms every state in the representation $(I_L,I_R)$
into the state in the representation $(I_R,I_L)$. We can
construct definite parity states only combining basis vectors from both
these irreducible representations. Hence it is only the direct sum of these
two representations
 
\begin{equation}
(I_L,I_R) \oplus (I_R, I_L), ~~~~~~ I_L \neq I_R,
\label{mult}
\end{equation}

\noindent
 that is invariant under parity. This reducible representation of the
chiral group is an irreducible representation of the larger
group, the parity-chiral group 

\begin{equation}
 SU(2)_L \times SU(2)_R \times C_i,
\label{gr}
\end{equation}

\noindent
where the group $C_i$ consists of two elements: identity
and inversion in 3-dimensional space.\footnote{
In the literature language is sometimes used in
a sloppy way and the representation (\ref{mult})
is referred to erroneously as an irreducible representation of the
chiral group.}
This symmetry group is
the symmetry of the QCD Lagrangian (neglecting quark masses),
 however only its
subgroup $SU(2)_I \times C_i$ survives in the broken symmetry mode.
The dimension of the representation (\ref{mult}) is

\begin{equation}
dim_{ (I_a,I_b) \oplus (I_b,I_a)} = 2(2I_a+1)(2I_b+1).
\label{dim}
\end{equation}
\\

 When we consider mesons
of isospin $I=0,1$, only three types of irreducible representations
of the parity-chiral group exist.\\

{\bf (i)~~~ (0,0).} Mesons in this representation must have isospin
$I=0$. At the same time $I_R=I_L=0$. This can be achieved when 
either there are no valence quarks in the meson\footnote{Hence glueballs must
be classified according to this representation \cite{G5}; with
no quark content this representation contains the state of only
one parity.}, or both
valence quark and antiquark are right or  left.
If we denote $R=(u_R,d_R)$  and $L=(u_L,d_L)$, then the basis
states of both parities can be written as

\begin{equation}
|(0,0); \pm; J \rangle = \frac{1}{\sqrt 2} (\bar R R \pm \bar L L)_J.
\label{00}
\end{equation} 

\noindent
Note that such a system can have spin $J \geq 1$. Indeed, valence
quark and antiquark in the state (\ref{00}) have definite
helicities, because generically helicity = +chirality for quarks and
helicity = -chirality for antiquarks. Hence the total spin projection
of the quark-antiquark system  onto the momentum direction of the quark
is $ \pm 1$. The parity transformation property of the quark-antiquark state
is then regulated by the total spin of the system \cite{LANDAU}

\begin{equation}
\hat P |(0,0); \pm; J \rangle = \pm (-1)^J  |(0,0); \pm; J \rangle.
\label{P00}
\end{equation} 
\\

{\bf (ii)~~~ (1/2,1/2).} In this case the quark must be right
and the antiquark must be left, and vice versa. These representations combine
states with I=0 and I=1, which must be of opposite parity.
The basis states
within the two distinct representations  (denoted as "a" and "b", respectively)
of this type are

\begin{equation}
|(1/2,1/2)_a; +;I=0; J \rangle = \frac{1}{\sqrt 2} (\bar R L + \bar L R)_J,
\label{I0+1}
\end{equation} 

\begin{equation}
|(1/2,1/2)_a; -;I=1; J \rangle = \frac{1}{\sqrt 2} (\bar R \vec \tau L -
\bar L \vec \tau R)_J,
\label{I0+2}
\end{equation} 

\noindent
and

\begin{equation}
|(1/2,1/2)_b; -;I=0; J \rangle = \frac{1}{\sqrt 2} (\bar R L - \bar L R)_J,
\label{I0-1}
\end{equation}

\begin{equation}
|(1/2,1/2)_b; +;I=1; J \rangle = \frac{1}{\sqrt 2} (\bar R \vec \tau L +
\bar L \vec \tau R)_J.
\label{I0-2}
\end{equation}

\noindent
In these expressions $\vec \tau$ are isospin Pauli matrices.
The parity of  every state in the representation is determined as

\begin{equation}
\hat P |(1/2,1/2); \pm; I; J \rangle = \pm (-1)^J |(1/2,1/2); 
\pm; I; J \rangle.
\label{P12}
\end{equation} 

The mesons in the representations of this type can have
any spin.
Note that the two distinct $(1/2,1/2)_a$ and $(1/2,1/2)_b$
 irreducible representations
of $SU(2)_L \times SU(2)_R$ form one irreducible representation
of $U(2)_L \times U(2)_R$.
\\

{\bf (iii)~~~ (0,1)$\oplus$(1,0).} The total isospin is 1 and
the quark and  antiquark must both be right or  left.
 This representation
is possible only for $J \geq 1$. The basis states are

\begin{equation}
|(0,1)+(1,0); \pm; J \rangle = \frac{1}{\sqrt 2} (\bar R \vec \tau R 
\pm \bar L  \vec \tau L)_J
\label{10}
\end{equation} 

\noindent
with  parities

\begin{equation}
\hat P |(0,1)+(1,0); \pm; J \rangle = \pm (-1)^J 
 |(0,1)+(1,0); \pm; J \rangle.
\label{P10}
\end{equation} 
\\

In the chirally restored regime the physical states must fill
out completely some or all of these representations.
We have to stress that the usual quantum numbers $I,J^{PC}$ are not
enough to specify the chiral representation for $J \geq 1$. 
It happens that some of
the physical particles with the given $I,J^{PC}$ belong to
one chiral representation (multiplet), while the other particles with
the same $I,J^{PC}$ belong to the other multiplet. Classification
of the particles according to $I,J^{PC}$ is simply not complete
in the chirally restored regime. This property will have very
important implications as far as the amount of the states
with  the given  $I,J^{PC}$ is
concerned.\\

In order to make this point clear, we will discuss some
of the examples. Consider first the mesons of spin $J=0$,
which are $ \pi,f_0,a_0$ and $\eta$ mesons with the $u,d$
quark content only.
The interpolating fields
are given as

\begin{equation}
 J_\pi(x)  = \bar q(x) \vec \tau \imath \gamma_5 q(x),
\label{pi}
\end{equation}

\begin{equation}
 J_{f_0}(x)  = \bar q(x)  q(x),
\label{f0}
\end{equation}

\begin{equation}
 J_{\eta}(x)  = \bar q(x)  \imath \gamma_5 q(x),
\label{pi}
\end{equation}

\begin{equation}
 J_{a_0}(x)  = \bar q(x) \vec \tau  q(x).
\label{pi}
\end{equation}

\noindent
These four currents belong to the
irreducible representation
of the 
$U(2)_L\times U(2)_R = SU(2)_L\times SU(2)_R \times U(1)_V\times U(1)_A$
group. It is instructive to see how these currents transform
under different subgroups of the group above.
\\

 The $SU(2)_L\times SU(2)_R$ transformations consist of vectorial
 and axial transformations in the isospin space (10). The axial
transformations mix the currents of opposite parity:

\begin{equation}
 J_\pi(x)  \leftrightarrow J_{f_0}(x) 
\label{pif0}
\end{equation} 

\noindent
as well as

\begin{equation}
 J_{a_0}(x)  \leftrightarrow J_{ \eta}(x).
\label{a0eta}
\end{equation} 

\noindent
The currents (\ref{pif0}) form the basis of the $(1/2,1/2)_a$
representation of the parity-chiral group, while the interpolators
(\ref{a0eta}) transform as $(1/2,1/2)_b$.\\

The $U(1)_A$ transformation  (4) mixes the currents
of the same isospin but opposite parity:

\begin{equation}
 J_\pi(x)  \leftrightarrow J_{a_0}(x) 
\label{pia0}
\end{equation} 

\noindent
as well as

\begin{equation}
 J_{f_0}(x)  \leftrightarrow J_{ \eta}(x).
\label{f0eta}
\end{equation}

\noindent
All four currents together belong to the  representation
$(1/2,1/2)_a \oplus (1/2,1/2)_b$ which
is an irreducible representation of the $U(2)_L\times U(2)_R $ group.\\

If the vacuum were invariant with respect to $U(2)_L\times U(2)_R $
transformations, then all four mesons, $\pi,f_0,a_0$ and $ \eta$
would be degenerate (as well as all their excited states). Once
the $U(1)_A$ symmetry is broken explicitly through
the axial anomaly, but the chiral $SU(2)_L\times SU(2)_R $ 
symmetry is still
intact in the vacuum, then the spectrum would consist of
degenerate $(\pi, f_0)$ and $(a_0,  \eta)$ pairs. If
in addition the chiral  $SU(2)_L\times SU(2)_R $ symmetry is
spontaneously broken 
in the vacuum, the degeneracy is also lifted in  the pairs
above and the pion becomes a (pseudo)Goldstone boson. Indeed,
the masses of the lowest mesons  are \cite{PDG}\footnote{The
$\eta$ meson mass given here was obtained by unmixing the
 $SU(3)$ flavor octet  and singlet states so it represents
the pure $\bar n n = (\bar u u + \bar d d )/\sqrt 2$ state,
see for details ref. \cite{G2}.}

 $$ m_\pi \simeq 140 MeV, ~m_{f_0} \simeq 400 - 1200 MeV,~
m_{a_0} \simeq 985 MeV ,~ m_{\eta } \simeq 782 MeV. $$

\noindent
This immediately shows that both $SU(2)_L\times SU(2)_R $ and
$U(1)_V \times U(1)_A$ are broken in the QCD vacuum
 to $SU(2)_I$ and $U(1)_V$, respectively.\\

If one looks at the upper part of the spectrum, then
one notices that
the four successive highly excited $\pi$ mesons
and the corresponding $\bar n n$ $f_0$ mesons 
form approximate chiral pairs \cite{G2}. This is well seen from the Fig. 5.
This pattern is a clear manifestation of the chiral
symmetry restoration. However, given the importance of
this statement these highly excited $\pi$ and $f_0$
mesons must be reconfirmed in other kind of experiments.\\

\begin{figure}
\centerline{
\psfig{file=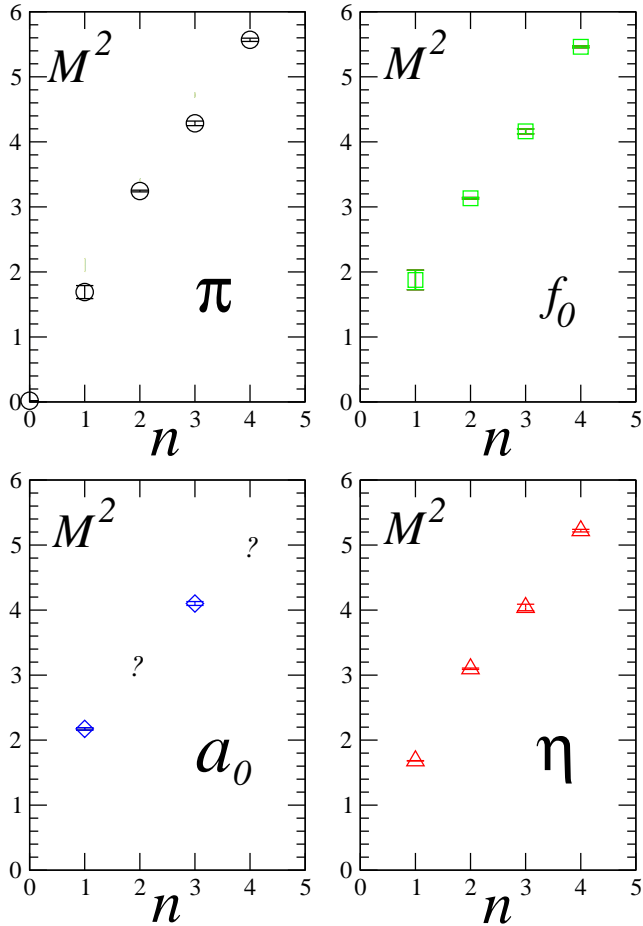,width=0.7\textwidth}
}
\caption{Radial Regge trajectories for the four successive
high-lying $J=0$ mesons.}
\end{figure}

A similar behaviour is observed from a comparison of the
$a_0$ and $\eta$ masses \cite{G2}. However, there are two
missing $a_0$ mesons which must be discovered in order to complete
all chiral multiplets. (Technically the identification of the
spinless states from the partial wave analysis is a rather
difficult task). There is a little doubt that these
missing $a_0$ mesons do exist. If one puts the four high-lying 
$\pi$, $\bar n n$ $f_0$, $a_0$ and  $\bar n n$ $\eta$ mesons on the 
{\it radial} Regge trajectories, see Fig. 9, one clearly notices
that the two missing $a_0$ mesons lie on the linear trajectory with
the same slope as all other mesons \cite{BUGG1,BUGG2}. 
If one reconstructs these
missing $a_0$ mesons according to this slope, then a pattern
of the $a_0 - \eta$ chiral partners appears, similar to the one
for the $\pi$ and $f_0$ mesons.\\

For the $J \geq 1$ mesons the classification is a bit
more complicated.
Consider $\rho(1,1^{--})$ mesons as example. Particles
of this kind can be created from the vacuum by the vector
current, $\bar \psi \gamma^\mu \vec \tau \psi$. Its chiral
partner is the axial vector current, 
$\bar \psi \gamma^\mu \gamma^5 \vec \tau \psi$, which creates
from the vacuum the axial vector mesons, $a_1(1,1^{++})$ . Both
these currents belong to the representation (0,1)+(1,0) and
have the  right-right $\pm$ left-left quark content. 
Clearly, in
the chirally restored regime the mesons created by these currents
must be degenerate level by level and fill out the (0,1)+(1,0)
representations. Hence, naively the amount of $\rho$ and $a_1$
mesons high in the spectrum should be equal.
This is not correct, however.   $\rho$-Mesons can be also created
from the vacuum by  other type(s) of  current(s),
$ \bar \psi \sigma^{0i} \vec \tau \psi$ (or by 
$\bar \psi \partial^\mu \vec \tau \psi$). These interpolators belong
to the (1/2,1/2) representation and have the 
left-right $\pm$ right-left quark content.
In the regime where chiral symmetry is strongly broken (as
in the low-lying states) the physical states are mixtures
of different representations. Hence these low-lying states are
well coupled to both (0,1)+(1,0) and (1/2,1/2) interpolators.
However, when chiral symmetry is  (approximately)
restored, then each physical
state must be strongly dominated by the given representation
and hence will  couple only to the interpolator which belongs
to the same representation.
This means that 
$\rho$-mesons created
by two distinct currents in the chirally restored regime represent
physically different particles.  The chiral partner of the
$ \bar \psi \sigma^{0i} \vec \tau \psi$ (or 
$\bar \psi \partial^\mu \vec \tau \psi$) current is
$ \varepsilon^{ijk} \bar \psi \sigma^{jk}  \psi$ (
$\bar \psi \gamma^5 \partial^\mu   \psi$, respectively)
\footnote{Chiral transformation properties of some 
interpolators can be found in ref. \cite{CJ}.}. The
latter interpolators create from the vacuum $h_1(0,1^{+-})$
states. Hence in the chirally restored regime, some of the
$\rho$-mesons must be degenerate with the $a_1$ mesons ((0,1)+(1,0)
multiplets), but the others
- with the $h_1$ mesons ((1/2,1/2) multiplets)\footnote{
Those $\rho(1,1^{--})$ and $\omega(0,1^{--})$ mesons which
belong to (1/2,1/2) cannot be seen in $e^+e^- \rightarrow hadrons$.}. 
Consequently, high in the spectra the combined amount
of $a_1$ and $h_1$ mesons must coincide with the amount of
$\rho$-mesons. This is a highly nontrivial prediction of chiral
symmetry.\\

Actually it is a very typical situation. Consider 
  $f_2(0,2^{++})$ mesons as another example. They can be interpolated
by the tensor field $\bar \psi \gamma^\mu \partial^\nu \psi$
(properly symmetrised, of course), which belongs to the (0,0)
representation. Their chiral partners are $\omega_2(0,2^{--})$
mesons, which are created by the 
$\bar \psi \gamma^5\gamma^\mu \partial^\nu \psi$ interpolator.
On the other hand  $f_2(0,2^{++})$ mesons can also be created from
the vacuum by the $\bar \psi \partial^\mu \partial^\nu \psi$
type of interpolator, which belongs to the (1/2,1/2) representation.
Its chiral partner is 
$\bar  \psi \gamma^5\partial^\mu \partial^\nu \vec \tau \psi$,
which creates $\pi_2(1,2^{-+})$ mesons. Hence in the chirally
restored regime we have to expect $\omega_2(0,2^{--})$ mesons
to be degenerate systematically with some of the $f_2(0,2^{++})$ mesons
((0,0) representations) while $\pi_2(1,2^{-+})$ mesons must
be degenerate with {\it other} $f_2(0,2^{++})$ mesons (forming
(1/2,1/2) multiplets). Hence the total number of $\omega_2(0,2^{--})$
and $\pi_2(1,2^{-+})$ mesons in the chirally restored regime
must coincide with the amount of $f_2(0,2^{++})$ mesons.\\

These examples can be generalized to mesons of any spin $J \geq 1$.
Those interpolators which contain only derivatives
$\bar \psi \partial^\mu \partial^\nu ...\psi$ ( 
$\bar  \psi \vec \tau \partial^\mu \partial^\nu ...\psi$)
have quantum numbers $I=0,P=(-1)^J,C=(-1)^J$ ($I=1,P=(-1)^J,C=(-1)^J$)
and transform as (1/2,1/2).
Their chiral partners are 
$\bar \psi \vec \tau  \gamma^5 \partial^\mu \partial^\nu ...\psi$
( $\bar  \psi \gamma^5\partial^\mu \partial^\nu ...\psi$, respectively)
with $I=1,P=(-1)^{J+1},C=(-1)^J$ ($I=0,P=(-1)^{J+1},C=(-1)^J$,
respectively). However, interpolators
with the same $I,J^{PC}$ can be also obtained with one $\gamma^\eta$
matrix instead    one of the  derivatives, $\partial^\eta$:
$\bar \psi \partial^\mu \partial^\nu ...\gamma^\eta...   \psi$ ( 
$\bar  \psi \vec \tau \partial^\mu \partial^\nu ... \gamma^\eta ...\psi$).
These
latter interpolators  belong  to (0,0)  ((0,1)+(1,0)) representation. 
Their chiral partners are
$\bar \psi \gamma^5 \partial^\mu \partial^\nu ...\gamma^\eta...   \psi$ ( 
$\bar  \psi \vec \tau  \gamma^5 \partial^\mu \partial^\nu ... 
\gamma^\eta ...\psi$) which have 
$I=0,P=(-1)^{J+1},C=(-1)^{J+1}$ ($I=1,P=(-1)^{J+1},C=(-1)^{J+1}$).
Hence in the chirally restored regime
the physical states created by these  different types of interpolators
 will belong to different
representations and will be distinct particles
while having the same $I,J^{PC}$. One needs to indicate
chiral representation in addition to  usual quantum numbers
$I,J^{PC}$ in order to uniquely  specify physical states of the $J \geq 1$
 mesons in the
chirally restored regime.\\

The available data for the $J=1,2,3$ mesons are systematized in 
Ref. \cite{G3}. Below we show the chiral patterns for the $J=2$
mesons, where the data set seems to be complete.\\

\begin{center}{\bf (0,0)}\\

{$\omega_2(0,2^{--})~~~~~~~~~~~~~~~~~~~~~~~f_2(0,2^{++})$}\\
\medskip
{$1975 \pm 20   ~~~~~~~~~~~~~~~~~~~~~~~~1934 \pm 20$}\\
{$2195 \pm 30   ~~~~~~~~~~~~~~~~~~~~~~~~2240 \pm 15$}\\

\bigskip
{\bf (1/2,1/2)}\\

{$\pi_2(1,2^{-+})~~~~~~~~~~~~~~~~~~~~~~~f_2(0,2^{++})$}\\
\medskip
{$2005 \pm 15   ~~~~~~~~~~~~~~~~~~~~~~~~2001 \pm 10$}\\
{$2245 \pm 60   ~~~~~~~~~~~~~~~~~~~~~~~~2293 \pm 13$}\\

\bigskip
{\bf (1/2,1/2)}\\

{$a_2(1,2^{++})~~~~~~~~~~~~~~~~~~~~~~~ \eta_2(0,2^{-+})$}\\
\medskip
{ $2030 \pm 20  ~~~~~~~~~~~~~~~~~~~~~~~~2030 ~\pm ~?$}\\
{ $2255 \pm 20 ~~~~~~~~~~~~~~~~~~~~~~~~2267 \pm 14$}\\

\bigskip
{\bf (0,1)+(1,0)}\\

{$a_2(1,2^{++})~~~~~~~~~~~~~~~~~~~~~~~\rho_2(1,2^{--})$}\\
\medskip
{ $1950^{+30}_{-70}~~~~~~~~~~~~~~~~~~~~~~~~~~~1940 \pm 40$}\\
{ $2175 \pm 40  ~~~~~~~~~~~~~~~~~~~~~~~~2225 \pm 35$}\\
\end{center}

\noindent
We see systematic patterns of chiral symmetry restoration. In
particular, the amount of $f_2(0,2^{++})$ mesons coincides with
the combined amount of $\omega_2(0,2^{--})$ and $\pi_2(1,2^{-+})$
states. Similarly,  number of $a_2(1,2^{++})$ states is
the same as number of $\eta_2(0,2^{-+})$ and $\rho_2(1,2^{--})$
together. All chiral multiplets are complete. While masses of
some of the states can and will be  corrected in the future experiments,
if  new  states might be discovered in this energy region in other
types of experiments, they
should be either $\bar s s$ states or glueballs.\\

The data sets for the $J=1$ and $J=3$ mesons are less complete
and there are a few missing states to be discovered \cite{G3}. Nevertheless,
these spectra also offer an impressive patterns of chiral symmetry.\\ 

It is important to see whether there are also signatures
of the $U(1)_A$ restoration. This 
can happen if two conditions are fulfilled \cite{CG1}: (i) unimportance
of the axial anomaly in excited states, (ii) chiral
$SU(2)_L \times SU(2)_R$ restoration (i.e. unimportance of the
quark condensates which break simultaneously both types
of symmetries in the vacuum state). 
Some evidence for the $U(1)_A$ restoration
has been reported in ref. \cite{G1} on the basis of $J=0$
data. Yet  missing $a_0$  states have
to be discovered to complete the $U(1)_A$ multiplets in the
$J=0$ spectra. In this section we will demonstrate that the
data on e.g. $J=2$ mesons present convincing evidence on $U(1)_A$
restoration.\\

First, we have to consider which mesonic states can be 
expected to be $U(1)_A$ partners. The $U(1)_A$ transformation
connects interpolators of the same isospin but opposite parity.
But not all such interpolators can be connected by the $U(1)_A$
transformation. For instance, the vector currents 
$\bar \psi \gamma^\mu \psi$
and $\bar \psi \vec \tau\gamma^\mu \psi$ are invariant under
$U(1)_A$. Similarly, the axial vector interpolators
 $\bar \psi \gamma^5 \gamma^\mu \psi$
and $\bar \psi \vec \tau \gamma^5 \gamma^\mu \psi$ are also invariant under
$U(1)_A$. Hence those interpolators (states) that are members
of the $(0,0)$ and $(0,1)+(1,0)$ representations of
$SU(2)_L \times SU(2)_R$ are invariant with respect to
$U(1)_A$. However,  interpolators (states) from the {\it distinct}
(1/2,1/2) representations which have the same isospin but
opposite parity  transform into each other under $U(1)_A$.
For example, $\bar \psi \psi \leftrightarrow \bar \psi \gamma^5 \psi$,
 $\bar \psi \vec \tau \psi \leftrightarrow \bar \psi \vec \tau \gamma^5 \psi$,
and those with derivatives: 
$\bar \psi \partial^\mu \psi \leftrightarrow \bar \psi \gamma^5  
\partial^\mu \psi$, $\bar \psi \vec \tau \partial^\mu \psi 
\leftrightarrow \bar \psi \vec \tau \gamma^5  \partial^\mu \psi$, etc.
If the corresponding states are systematically degenerate, then
it is a signal that $U(1)_A$ is restored. In what follows we show
that it is indeed the case.

\begin{center}

{$f_2(0,2^{++})~~~~~~~~~~~~~~~~~~~~~~~\eta_2(0,2^{-+})$}\\
\medskip
{$2001 \pm 10  ~~~~~~~~~~~~~~~~~~~~~~~~2030 ~\pm ~?$}\\
{$2293 \pm 13   ~~~~~~~~~~~~~~~~~~~~~~~~2267 \pm 14$}\\

\bigskip

{$\pi_2(1,2^{-+})~~~~~~~~~~~~~~~~~~~~~~~a_2 (1,2^{++})$}\\
\medskip
{$2005 \pm 15  ~~~~~~~~~~~~~~~~~~~~~~~~ 2030 \pm 20$}\\
{$2245 \pm 60 ~~~~~~~~~~~~~~~~~~~~~~~~ 2255 \pm 20 $}\\
\end{center}

 We see clear approximate doublets of $U(1)_A$ restoration. Hence
two distinct (1/2,1/2) multiplets of $SU(2)_L \times SU(2)_R$
can be combined into one multiplet of $U(2)_L \times U(2)_R$.
So we conclude that the whole chiral symmetry  of the
QCD Lagrangian $U(2)_L \times U(2)_R$ gets approximately restored high in the hadron spectrum.\\

It is useful to quantify the effect of chiral symmetry breaking
(restoration). An obvious parameter that characterises effects
of chiral symmetry breaking is a relative mass splitting within
the chiral multiplet. Let us define the {\it chiral asymmetry} as

\begin{equation}
\chi = \frac{|M_1 - M_2|}{(M_1+M_2)},
\label{chir}
\end{equation}

\noindent
where $M_1$ and $M_2$ are masses of particles within the
same multiplet. This parameter gives a quantitative measure
of chiral symmetry breaking at the leading (linear) order
and has the interpretation of the part
 of the hadron mass  due to chiral
symmetry breaking.\\

For the low-lying states the chiral asymmetry is typically
0.2 - 0.6 which can be seen e.g. from a comparison of the $\rho(770)$
and $a_1(1260)$ or the  $\rho(770)$ and $h_1(1170)$ masses. If 
the chiral asymmetry is large as above, then it makes no
sense to assign a given hadron to the chiral multiplet
since its wave function is a strong mixture of different
representations and we have to expect also large
{\it nonlinear} symmetry breaking effects. However, at meson 
masses about 2 GeV
the chiral asymmetry is typically within 0.01
and in this case the hadrons can be believed to be  members
of  multiplets with a tiny admixture of other
representations. Unfortunately there are no systematic data
on mesons below 1.9 GeV and hence it is difficult to estimate
the chiral asymmetry as a function of mass ($\sqrt s$). Such
a function would be crucially important for a further progress
of the theory. So a systematic experimental study of hadron spectra is
difficult to overestimate. However, thanks to the $0^{++}$
glueball search for the last 20 years, there are such data
for $\pi$ and $f_0$ states, as can be seen from Fig. 5
(for details we refer to \cite{G2,G5}). According to these data
we can reconstruct $\chi(\sqrt s \sim 1.3 GeV) \sim 0.03 \div 0.1$,
$\chi(\sqrt s \sim 1.8 GeV) \sim 0.008$, 
$\chi(\sqrt s \sim 2.3 GeV) \sim 0.005$.
We have to also  stress that there is no reason to expect
the chiral asymmetry to be a universal function for all
hadron channels. Hadrons with different quantum numbers
feel chiral symmetry breaking effects differently, as can
be deduced from the operator product expansions of two-point
functions for different currents.
 A task of the theory is to derive these
chiral asymmetries microscopically.\\

\section{Chiral multiplets of excited baryons}

Now we will consider chiral multiplets of excited baryons \cite{CG1,CG2}.
The nucleon or delta states have a half integral isospin. Then
such a multiplet cannot be an irreducible representation of the
chiral group $(I_L,I_R)$ with $I_L = I_R$, because in this case
the total isospin can only be integral. Hence the minimal possible
representation that is invariant under parity transformation is
the one of (\ref{mult}).
Empirically, there are no known baryon resonances within the two
light flavors sector which have an isospin greater than 3/2.
Thus we have a constraint from the data that
if chiral symmetry is effectively restored for very highly
excited baryons, the only possible representations for the
 observed baryons have $I_L + I_R \le 3/2$, {\it i.e.} the
 only possible representations are
$(1/2,0) \oplus (0,1/2)$, $(1/2,1) \oplus (1,1/2)$
 and $(3/2,0) \oplus (0,3/2)$. Since chiral symmetry and
parity do not constrain the possible spins of the states
these multiplets can correspond to states of any fixed spin.\\

The same classification can actually be obtained assuming that
chiral properties of excited baryons are determined by three
massless valence quarks which have a definite chirality. Indeed
the one quark field transforms as

\begin{equation}
q \sim \left (\frac{1}{2},0 \right ) \oplus \left (0,\frac{1}{2} \right).
\label{qq}
\end{equation}
Then all possible representations for the three-quark baryons
in the chirally restored phase can
be obtained as a direct product of three "fundamental" representations
(\ref{qq}). Using the standard isospin coupling rules
separately for the left and right quark components, one
easily obtains a decomposition of this direct product

$$
\left[\left (\frac{1}{2},0 \right ) 
\oplus \left (0,\frac{1}{2} \right )\right]^3 =
\left[\left (\frac{3}{2},0 \right ) 
\oplus \left (0,\frac{3}{2} \right )\right]
$$

\begin{equation}
+ 3\left[ \left (1,\frac{1}{2}\right ) 
\oplus \left (\frac{1}{2},1\right )\right]
+ 3\left[\left (0,\frac{1}{2} \right ) 
\oplus \left (\frac{1}{2},0\right )\right]
+ 2\left[\left (\frac{1}{2},0 \right ) 
\oplus \left (0,\frac{1}{2} \right )\right]
.
\label{dec}
\end{equation}

\noindent
The last two representations in the expansion above
are identical group-theoretically, so they can be combined
with the common multiplicity factor 5.
Thus, according to the simple-minded model above,
baryons in the chirally
restored regime will belong to one of the
following representations:
\begin{equation}
\left(\frac{1}{2},0 \right) \oplus \left(0,\frac{1}{2} \right);~
\left(\frac{3}{2},0 \right) \oplus \left(0,\frac{3}{2} \right);~
\left(\frac{1}{2},1 \right) \oplus \left(1,\frac{1}{2} \right).
\label{list}
\end{equation}
\\

The $(1/2,0) \oplus (0,1/2)$ multiplets contain only isospin
1/2 states and hence correspond to parity doublets of nucleon
states (of any fixed spin).\footnote{If one distinguishes
nucleon states with different electric charge, i.e. different
isospin projection, then this ``doublet'' is actually a quartet.}
Similarly, $(3/2,0) \oplus (0,3/2)$
 multiplets contain only isospin 3/2 states and hence correspond
to parity doublets of $\Delta$ states (of any fixed spin).\footnote{
Again, keeping in mind different charge states of delta
resonance it is actually an octet.}
However, $(1/2,1) \oplus (1,1/2)$ multiplets contain both
isospin 1/2 and isospin 3/2 states and hence correspond to
 multiplets containing both nucleon and $\Delta$ states of
both parities and any fixed spin.\footnote{This representation
is a 12-plet once we distinguish between different charge states.}\\

Summarizing, the phenomenological consequence of the effective
restoration of chiral symmetry
high in $N$ and $\Delta$ spectra is that the baryon states
will fill out  the irreducible
representations of the parity-chiral group (\ref{gr}).
If $(1/2,0) \oplus (0,1/2)$ and $(3/2,0) \oplus (0,3/2)$
multiplets were realized in nature, then the spectra of highly excited
nucleons and deltas would consist of parity doublets. However,
the energy of the parity doublet with  given spin in
the nucleon spectrum {\it a-priori} would not be degenerate with the
the doublet with the same spin in the delta spectrum;
these doublets would belong to different
representations of eq.~(\ref{gr}), {\it i.e.} to distinct
multiplets and their energies
are not related.   On the other hand,
if $(1/2,1) \oplus (1,1/2)$ were realized, then the highly
lying states in $N$ and $\Delta$ spectrum 
would have a $N$ parity doublet and a $\Delta$
parity doublet with the same spin and which are degenerate in mass.
In either of cases the highly lying spectrum  must systematically
consist of parity doublets.\\

If one looks carefully at the nucleon spectrum, see Fig. 4, and the 
delta spectrum
one notices that the systematic parity doubling in the nucleon
spectrum appears at  masses of 1.7 GeV and above, while the 
parity doublets in the delta spectrum insist at  masses of 
1.9 GeV.\footnote{This means that the parity doubling in both cases
is seen at approximately the same excitation energy with respect
to the corresponding ground state.} This fact implies that at
least those nucleon doublets that are seen at $\sim 1.7$GeV belong
to $(1/2,0) \oplus (0,1/2)$ representation. Below we show
doublets of different spin in the energy range of 1.9 GeV and higher:

$${\bf J=\frac{1}{2}:}
 ~N^+(2100)~(*),~N^-(2090)~(*),~\Delta^+(1910)~~~~,~\Delta^-(1900) (**);$$

$$ {\bf J=\frac{3}{2}:}
 ~N^+(1900) (**),~N^-(2080) (**),~\Delta^+(1920)~~~~,~\Delta^-(1940)~(*);$$

$${\bf J=\frac{5}{2}:}
 ~N^+(2000) (**),~~N^-(2200) (**),~ \Delta^+(1905)~~~~,~\Delta^-(1930)~~~~;$$

$${\bf J=\frac{7}{2}:}
 ~N^+(1990) (**),~ N^-(2190)~~~~,~\Delta^+(1950)~~~~,~\Delta^-(2200)~(*);$$

$${\bf J=\frac{9}{2}:}
 ~N^+(2220)~~~,~N^-(2250)~~~~,~\Delta^+(2300) (**),~\Delta^-(2400)(**);$$

$${\bf J=\frac{11}{2}:}
 ~~~~~~~~~?~~~~~~~~,~N^-(2600)~~~~,~\Delta^+(2420)~~~~,~~~~~~~~?~~~~~~~~;$$

$${\bf J=\frac{13}{2}:}
 ~N^+(2700) (**), ~~~~~~~?~~~~~~~~,~~~~~~~~?~~~~~~~~,~\Delta^-(2750)(**);$$

$${\bf J=\frac{15}{2}:}
 ~~~~~~~~?~~~~~~~~,  ~~~~~~~~?~~~~~~~~,~\Delta^+(2950) (**),~~~~~~~~?~~~~~~~~.$$

If approximate mass degeneracy between $N$ and $\Delta$ doublets 
at $M \geq 1.9$ GeV is accidental, then the baryons in this mass
region are
organized according to 
$(1/2,0) \oplus (0,1/2)$ for $N$ and
$(3/2,0) \oplus (0,3/2)$ for $\Delta$ parity-chiral doublets. If not,
then the high lying spectrum forms $(1/2,1) \oplus (1,1/2)$ multiplets.
It can also be possible that in the narrow energy interval
more than one parity doublet in the nucleon and delta spectra
is found for a given spin. This would then mean that different
doublets  belong to different parity-chiral multiplets.
Systematic experimental exploration of the high-lying states
is required in order to assign unambiguously baryons to
the multiplets.\\

\section{Can simple potential models explain parity doubling?}

Before discussing a model for highly excited hadrons that
is compatible with the chiral symmetry restoration and parity
doubling it is useful to answer a question whether the
potential models like the traditional constituent quark model
can explain it. Consider first mesons. Within the potential model
the mesons are considered to be systems of two constituent
quarks which interact via linear confinemet potential plus
some perturbation from the one gluon exchange \cite{GI} or instanton
induced interaction \cite{METSCH}. Within the potential
description the parity of the state is unambiguously prescribed
by the relative orbital angular momentum $L$ of the constituent
quarks. For example, all the states on the radial pion Regge
tragectory, see Fig. 9, are $^1S_0$ $\bar Q Q$ states, while the
members of the $f_0$ trajectory are the $^3P_0$ states. Hence the
centrifugal repulsion for the states of opposite parity is
different. Then it is clear that such a model cannot explain a
systematic approximate degeneracy of the states of opposite parity.
A fine tuning of the perturbation can in principle provide
an {\it accidental} degeneracy of some of the states, but then
there will be no one-to-one pairing and degeneracy for the
other states. As a consequence the potential models of 
mesons cannot accomodate a lot of experimentally
observed highly excited mesons. For example, while the parameters
within the model of ref. \cite{GI} are fitted to describe the two
lowest pion states and it still can accomodate the third radial
state of the pion, it does not predict at all the existence
of $\pi(2070)$ and $\pi(2360)$; the fourth and the fifth radial
states of the pion do not appear in this picture up to 2.4 GeV.
A similar situation occurs also in other channels. The failure
of the potential description is inherently related to the fact
that it cannot incorporate chiral symmetry restoration as a matter
of principle. The latter phenomenon is intrinsically a
relativistic phenomenon which is a consequence of the fact that
the ultrarelativistic valence quarks in the highly excited hadrons
must necessarily be chiral (i.e. they have definite helicity and
chirality). It is a generic property of the ultrarelativistic
fermions which cannot be simulated within the $^{2S+1}L_J$ type
potential description.\\

If one uses instead a relativistic description
within the Dirac or Bethe-Salpeter equations frameworks, then the
parity doubling and chiral symmetry restoration is incompatible with
the Lorentz scalar potential which is often used to
simulate confinement. The reason is that the Lorentz scalar potential
manifestly breaks chiral symmetry and is equivalent to introduction
of some effective mass which increases with the excitation and 
size of hadrons. With the Lorentz vector confining potential
and assuming that there is no constituent mass of quarks
one can obtain parity doubling \cite{Y}.\\

A few comments about the parity doubling within the potential
models that attempt to describe the highly lying baryons
are in order. The models that rely on  confinement potential
cannot explain an appearance of the systematic parity doublets.
This is apparent for the harmonic confinement. The parity of
the state is determined by the number $N$  of the harmonic
excitation quanta in the 3q state. The ground states (N=0) are
of positive parity, all baryons from the $N=1$ band are of negative
parity, baryons from the $N=2$ band have a positive parity irrespective
of their angular momentum, etc. However, the number of states in
the given band rapidly increases with $N$. This means that such a
model cannot provide an equal amount of positive and negative
parity states, which is necessary for parity doubling, 
irrespective of other residual interactions between
quarks in such a model. Similar problem persists with the
linear confinement in the 3q system.\\ 

While all vacancies from the $N=0$ and $N=1$ bands are filled
in  nature, such a model, extrapolated to the N=3 and higher bands 
predicts
a very big amount of states, which are not observed (the so-called
missing resonance problem). The chiral restoration transition takes
place at excitation energies typical for the highest states in the
$N=2$ band and in the $N=3$ bands. 
If correct, it would mean that
description of  baryons in this transition region 
 in terms of constituent quarks becomes inappropriate.\\

The model that relies on the pure color Coulomb interaction between
quarks also cannot provide the systematical parity doubling. While
it gives an equal amount of the positive and negative parity
single quark states in the $n=2,4,...$ bands (e.g. $2s-2p$, or
$4s-4p,~ 4d-4f$), the number of the positive parity states
is always bigger in the $n=1,3,5,...$ bands.\\

\section{Chiral symmetry restoration and the string (flux tube) picture}

A question arises what is a microscopical mechanism of
chiral symmetry restoration in excited hadrons and what is
a relevant physical picture? We have already mentioned before
that a possible scenario is related to the fact that at large
space-like momenta  the dynamical (constituent)
mass of quarks must vanish.
If in the highly excited hadrons the momenta of valence quarks
are indeed large, then the effects of spontaneous breaking of
chiral symmetry should be irrelevant in such hadrons \cite{G1,swanson}.\\ 

Here we will discuss a possible fundamental origin for this
phenomenon. We will show below that both chiral and $U(1)_A$
restorations can be anticipated as a direct consequence of the
semiclasical regime in the highly excited hadrons.\\

At large $n$ (radial quantum number) or at large angular
momentum $L$ we know that in quantum systems the {\it semiclassical}
approximation (WKB) {\it must} work. Physically this approximation
applies in these cases because the de Broglie wavelength of
particles in the system is small in comparison with the
scale that characterizes the given problem. In such a system
as a hadron the scale is given by the hadron size while the
wavelength of valence quarks is given by their momenta. Once
we go high in the spectrum the size of hadrons increases as well as
 the typical momentum of valence quarks.
This is why a highly excited hadron  can be described semiclassically
in terms of the underlying quark  degrees of freedom.\\

A physical content of the semiclassical approximation is
most transparently given by the path integral. The contribution
of the given path to the path integral is regulated by the
action $S(q)$ along the  path $q(x,t)$

\begin{equation}
\sim e^{iS(q)/\hbar}.
\label{path}
\end{equation}

\noindent
The semiclassical approximation  applies when $S(q) \gg \hbar$.
In this case the whole amplitude (path integral) is dominated by
the classical path $q_{cl}$ (stationary point) and those paths that are infinitesimally
close to the classical path. All other paths that differ from
the classical one by an appreciable amount  do
not contribute. These latter paths would represent the quantum fluctuation
effects. In other words, in the semiclassical case the quantum
fluctuations effects are strongly suppressed and vanish asymptotically.\\

The $U(1)_A$ symmetry of the QCD Lagrangian is broken only due
to the quantum fluctuations of the fermions. The $SU(2)_R \times SU(2)_L$
spontaneous (dynamical) breaking is also pure quantum effect
and is based upon quantum fluctuations. To see the latter we
remind the reader that most generally the chiral symmetry breaking
(i.e.the dynamical quark mass generation) is formulated via the
Schwinger-Dyson  (gap) equation. It is not yet clear at all
which specific gluonic interactions are the most important ones as a kernel
of the Schwinger-Dyson equation (e.g. instantons \footnote{
The instanton itself is a semiclassical gluon field configuration.
But chiral and $U(1)_A$ symmetry breakings by instantons is
a quark field quantum fluctuations process. This is because
these breakings are due to chiral quark pair creation from
the vacuum by the instanton.}
, or 
gluonic exchanges, or perhaps  other gluonic
interactions, or a combination of different interactions).
But in any case the quantum fluctuations effects of the quark
 fields are
very strong in the low-lying hadrons and induce both 
chiral and $U(1)_A$ breakings. As a consequence we do not
observe any chiral or $U(1)_A$ multiplets low in the spectrum.
However,  if the quantum fluctuations effects
are relatively suppressed , then the 
dynamical mass of quarks must vanish as well as the effects of the $U(1)_A$
anomaly.\\

We have just mentioned that in a bound state quantum system with
large enough $n$ or  $L$ the effects of quantum fluctuations must be
suppressed and vanish asymptotically.\footnote{That the quantum 
fluctuations effects vanish in the quantum bound state 
systems at large $n$ or $L$ 
is well known e.g. from the Lamb shift. The Lamb shift is a result
of the radiative corrections (which represent effects
of quantum fluctuations of electron and electromagnetic fields)
 and vanishes as $1/n^3$,
and also very fast with increasing $L$. As a consequence high in the
hydrogen spectrum  the symmetry of the classical Coulomb
potential gets restored.}
Hence at large hadron masses (i.e. with either large $n$ or
large $L$) we should anticipate symmetries of the classical
QCD Lagrangian. Then it follows 
that in such systems both
the chiral and $U(1)_A$ symmetries must be restored. This is precisely what we see
phenomenologically. In the nucleon spectrum the doubling
appears either at large $n$ excitations of baryons with
the given small spin or in resonances of large spin. Similar
features persist in the delta spectrum. In the meson spectrum
the doubling is obvious for large $n$ excitations of small
spin mesons  and there are signs of doubling of large spin
mesons (the data are, however, sparse). It would be certainly
interesting and important to observe systematically multiplets
of parity-chiral and parity-$U(1)_A$ groups  (or, sometimes, when
the chiral and $U(1)_A$ transformations connect {\it different}
hadrons, the multiplets of the
$U(2)_L \times U(2)_R$ group). The high-lying hadron spectra
must be systematically explored.
\\

The strength of the argument given above is that it is very
general. Its weakness is that we
cannot say anything concrete about microscopical mechanisms of
how all this happens. For that one needs a detailed microscopical
understanding of dynamics in QCD, which is both challenging and
very difficult task. But even though we do not know 
how microscopically all this happens, we can anticipate that in highly excited
hadrons we must observe symmetries of the classical QCD Lagrangian.
The only basis for this statement is that in such hadrons a semiclassical
description is correct. \\

As a consequence, in highly excited hadrons the valence quark
motion has to be described semiclassically and at the same time
their chirality (helicity) must be fixed. Also the gluonic
field should be described semiclassically. All this gives an
increasing support for a string picture \cite{NAMBU}
of highly excited hadrons. Indeed, if one assumes that
the quarks at the ends of the string have definite
chirality, see Fig. 10, then all hadrons will appear necessarily
in chiral multiplets \cite{G4}. The latter hypothesis is very
natural and is well compatible with the Nambu string picture.
The ends of the string in the Nambu picture move with the
velocity of light. Then, (it is an extention of the Nambu model)
the quarks at the ends of the string must have definite chirality.
In this way one is able to explain at the same time both
Regge trajectories and parity doubling.\\

\begin{figure}
\centerline{\psfig{file=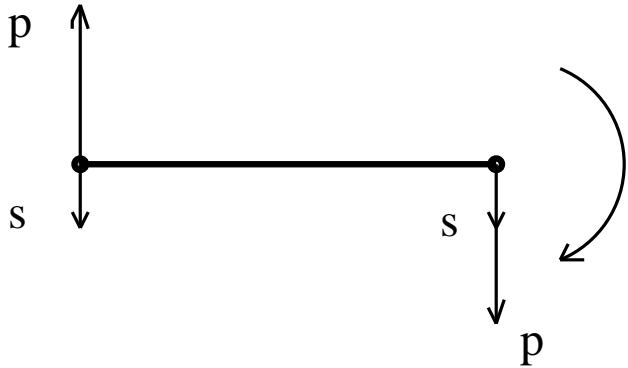,width=0.5\textwidth,angle=-90}}
\caption{ Rotating string with the right and the left quarks at the ends.}
\end{figure}

One arrives at the following situation: (i) the hadrons with the
different chiral configurations of the quarks at the
ends of the string which belong to the same parity-chiral
multiplet
and that belong to the same intrinsic quantum state of the string
must be degenerate; (ii) the total parity of the hadron is determined
by the product of parity of the string in the given quantum state
and the parity
of the specific parity-chiral configuration of the quarks at the
ends of the string. There is no analogy to this situation in the
nonrelativistic physics where  parity is only determined by the 
orbital motion of particles.
Thus one sees that for every intrinsic quantum state of the string
there necessarily appears parity doubling of the states
with the same total angular momentum.\\

The spin-orbit operator $\vec \sigma \cdot \vec L$ does not
commute with the helicity operator $\vec \sigma \cdot \vec \nabla$.
Hence the spin-orbit interaction of quarks with the fixed
chirality or helicity is absent. In particular, this is also true
for the spin-orbit force due to the Thomas precession

\begin{equation}
U_T = -\vec \sigma \cdot \vec \omega_T \sim \vec \sigma
 \cdot [\vec v, \vec a]
\sim \vec v \cdot [\vec v, \vec a] =0,
\label{T}
\end{equation}

\noindent
where $U_T$ is the energy of the interaction and
$\vec \omega_T$, $\vec v$ and $\vec a$ are the angular frequency
of Thomas precession, velocity of the quark and its acceleration, respectively.\\

The absense of the spin-orbit force in the chirally restored regime
is a very welcome feature because it is a well-known empirical fact
that the spin-orbit force is either vanishing or very small
in the spectroscopy in the $u,d$ sector \cite{S}. This fact is difficult,
if impossible, to explain within the potential constituent quark models.\\

In addition, for the rotating string

\begin{equation}
\vec \sigma (i) \cdot \vec R(i)=0,
\label{T1}
\end{equation}

\begin{equation}
 \vec \sigma (i) \cdot \vec R(j)=0,
\label{T2}
\end{equation}

\noindent
where the indices $i,j$ label different quarks and $ \vec R$  
is the radius-vector of the
given quark in the center-of-mass frame.
 The relations above
immediately imply that the possible tensor interactions
of quarks related to the string dynamics should
be absent, once the chiral symmetry is restored.\\

\section{Conclusions}

We have demonstrated in these lectures that the chiral symmetry of
QCD is crucially important to understand physics of hadrons in
the $u,d$ (and possibly in the $u,d,s$) sector. The low-lying hadrons
are mostly driven by the spontaneous breaking of chiral symmetry.
This breaking determines the physics and effective degrees of
freedom in the low-lying hadrons. For example, it is SBCS which
sheds a light on the meaning of the constituent quarks. The latter
ones are quasiparticles and appear due to coupling of the valence
quarks to the quark condensates of the vacuum. The pion as Nambu-Goldstone
boson represents a relativistic bound state of quasiparticles $Q$ and
$\bar Q$ and is a highly collective state in terms of original
bare quarks. A strength of the residual interaction between the
quasiparticles in the pion is dictated by chiral symmetry and is such
that it exactly compensates the mass of the constituent quarks so the pion
becomes massless in the chiral limit. In the low-lying baryons the
physics at low momenta is mostly dictated by the coupling of  constituent
quarks and Goldstone bosons. Then a crucially important residual
interaction between the constituent quarks in the low-lying baryons
is mediated by the pion field, which is of the flavor- and
spin-exchange nature.\\

However, this physics is relevant only to the low-lying hadrons. In
the high-lying hadrons the chiral symmetry is restored, which is
referred to as effective chiral symmetry restoration or chiral
symmetry restoration of the second kind. A direct manifestation of the
latter phenomenon is a systematical appearance of the approximate chiral
multiplets of the high-lying hadrons.
The essence of the present phenomenon is that the quark 
condensates which break chiral symmetry in the vacuum state (and
hence in the low-lying excitations over vacuum) become simply
irrelevant (unimportant) for the physics of the highly excited states.
The physics here is such as if there were no chiral symmetry
breaking in the vacuum. The valence quarks simply decouple from
the quark condensates and consequently the notion of the constituent
quarks with dynamical mass induced by chiral symmetry breaking
becomes irrelevant in highly excited hadrons.
Instead,
the string picture with quarks of  definite chirality at the
end points of the string should be invoked.
In  recent lattice calculations DeGrand has demonstrated that
indeed in the highly excited mesons valence quarks decouple
from the low-lying eigenmodes of the Dirac operator (which determine
the quark condensate via  Banks-Casher relation) and so decouple
from the quark condensate of the QCD vacuum \cite{degrand}.\\

Hence  physics of the high-lying hadrons is mostly physics
of confinement acting between the light quarks. Their very small current
mass strongly distinguishes this physics from the
physics of the havy quarkonium, where chiral symmetry is irrelevant
 and the string (flux tube) can be approximated as a static potential
acting between the slowly moving heavy quarks. In the light
hadrons in contrast the valence quarks are ultrarelativistic and their 
fermion
nature requires them to have a definite chirality. Hence the
high-lying hadrons in the $u,d$ sector  open a door to the regime of
dynamical strings with chiral quarks at the ends. Clearly a
systematic experimental exploration of the high-lying hadrons is 
required which is an interesting and important task and which should
be of highest priority at the existing accelerators and at the
forthcoming ones like PANDA at GSI and JPARC.\\

\section{Acknowledgement}
I am thankful to the organizers of the Zakopane school for
their kind hospitality.
The work was supported by the FWF project P16823-N08 of
the Austrian Science Fund. The author is indebted to C.B.
Lang for a careful reading of the manuscript.

\end{document}